\documentclass{iau}
\usepackage{graphicx}
 
\title{Models of AGN feedback}

\author[F.Combes]{Fran\c coise Combes$^1$}

\affiliation{$^1$Observatoire de Paris \\ 61 Av. de l'Observatoire
F-75 014 Paris, France\\ email: {\tt francoise.combes@obspm.fr}}

\pubyear{2014}
\volume{309}
\jname{Galaxies in 3D across the Universe}
\editors{B. L. Ziegler, F. Combes, H. Dannerbauer, M. Verdugo, eds.}

\begin{document}

\maketitle

\begin{abstract}
The physical processes responsible of sweeping up the
surrounding gas in the host galaxy of an AGN, and able
in some circumstances to expel it from the galaxy,
are not yet well known. The various mechanisms are briefly
reviewed: quasar or radio modes, either momentum-conserving outflows,
energy-conserving outflows, or intermediate. They are 
 confronted to observations, to know whether they can explain
the M-sigma relation, quench the star formation or whether they can also provide some
positive feedback and how the black hole accretion history is related 
to that of star formation.
\end{abstract}

\firstsection
\section{Introduction}

AGN feedback is invoked to prevent massive galaxies to form in too high abundance,
and to understand why the fraction of baryons in galaxies is very low, 
and decreasing with the galaxy mass (e.g. Behroozi et al. 2013).

It is clear that the growth of massive black holes in the center of galaxies
releases enough energy to have a large impact on the galaxy host, if this energy
is efficiently coupled to the matter.  Assuming the famous relation 
between black hole and bulge mass (e.g. G\"ultekin et al. 2009),
M$_{BH}$=1-2 10$^{-3}$ M$_{bul}$, and the radiative efficiency of accretion onto 
a black hole of 10\%,  E$_{BH}$ =0.1M$_{BH}$c$^2$, the binding energy of the bulge 
is E$_{bul}\sim$ M$_{bul} \sigma^2$, with $\sigma$ its velocity dispersion.
Typically, the ratio of these two energies is
 E$_{BH}$/E$_{bul}  >$ 100. There is therefore no problem in global energy,
however, this energy could be radiated away through an elongated cavity
perpendicular to the galaxy plane, and the true impact on the galaxy is unknown.

One can distinguish two main modes for AGN feedback
(see e.g. the recent review by Fabian, 2012):

-- the {\bf quasar mode}, through radiative processes or winds,
when the AGN luminosity is high, close to the Eddington limit.
This is the case for young QSOs, at high redshift. The Eddington
limit is 
L$_{Edd}$ = 4$\pi$GM$_{BH}$m$_p$c/$\sigma_T$, with $\sigma_T$
being the Thomson scattering cross-section between charged particles and photons. 
 When writing the equilibrium between the gravity and radiative pressure
on the infalling ionized gas, combined with the Virial relation, this gives
M$_{BH}\sim$f $\sigma_T \sigma^4$,  with f the gas fraction, which is quite
close to the M$_{BH}-\sigma$ relation. When applying the same considerations
at larger radii, this time for neutral gas, submitted to 
radiation pressure on dust, with the cross-section $\sigma_d$, this leads
a limitation on the mass than can be accumulated on the bulge
and the corresponding masses are in the ratio
M$_{bul}$/M$_{BH}$=$\sigma_d/\sigma_T\sim$1000,  which is
surprisingly close to the observed ratio.

-- the {\bf radio mode}, or kinetic mode with radio jets,
when the AGN luminosity is low, L $<$ 0.01 L$_{Edd}$, 
corresponding typically to low z, massive galaxies, like
the local radio ellipticals.
The process is then not destructive, and keeps a balance between cooling
and heating. This is the mode occuring in cool core clusters, from
the AGN of the central bright galaxy.
It is also associated to low-luminosity AGN in Seyferts.
It might be combined with a radiatively inefficient flow ADAF
(Advection dominated Accretion Flow).

The best evidence of AGN feedback is found in cool core clusters of galaxies,
where the active nucleus in the central galaxy moderates the cooling flow,
through its radio jet, creating bubbles in the hot gas.  This can have an impact
up to radii larger than the central galaxy, at 100kpc scales (McNamara et al. 2009).  About 
70 percent and possibly 90 percent of clusters have a cool core (Edge et al. 1992).
The clusters that are not cooling flows are unrelaxed and most probably
recent mergers of sub-clusters.

In the following, we will detail the AGN feedback moderating cooling flows,
describe recent results on molecular outflows, driven by starbursts and AGN,
and the nature of their mechanisms  and energetics. We finish by considerations on
the modes of quenching.

\section{AGN feedback moderating gas flow in cool core clusters}

It is frequent in cool core clusters to see the extended impact of the
central radio jet, under bubbles and cavities imprinted on the hot X-ray gas,
and cold filaments and streaks observed with H$\alpha$ and CO emission.
In the prototypical Perseus cluster, a large network of cold molecular  gas filaments
has been observed (Salom\'e et al. 2006). Recently Canning et al. (2014)
have identified a population of very young (a few Myr), compact star clusters,
over kiloparsecs scales, related to the network of filaments.
These stars could then enrich the stellar halo of the central galaxy NGC 1275.

Observations suggest that the filament network is a consequence of
both inflow and outflow. Some of the hot gas cools and fuels the central AGN,
which jets entrain the molecular gas in the central galaxy coming from previous cooling
episodes. The gas uplifted has been metal enriched in the galaxy before
mixing with the cooling gas,  and therefore can radiate in CO. 
The bubbles create inhomogeneities and further cooling. The latter is not only
occuring in the center of the cluster, but far away, up to 50 kpc, due
to the inhomogeneous density and compressions there.
The observed velocity of the gas in filaments is much lower than free-fall
(Salom\'e et al. 2008).
Numerical simulations have reproduced this process, with
buoyant bubbles triggering compression and gas cooling at the surface
of cavities, mixing with the cold gas dragged upwards
(Revaz et al. 2007).

\begin{figure}[ht]
\centering
\includegraphics[width=0.95\columnwidth]{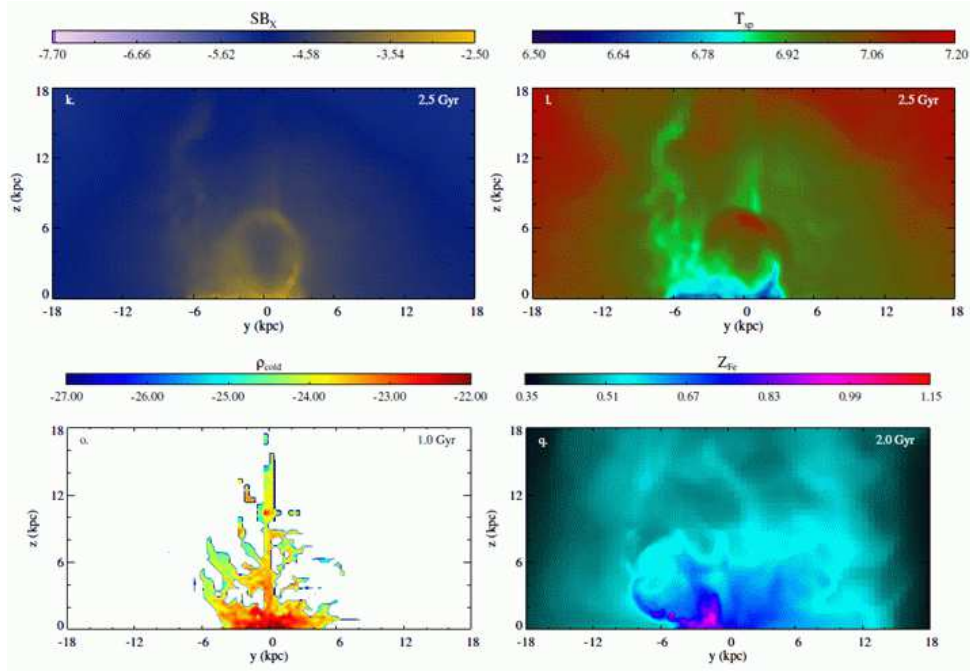} 
\caption{Simulation of AGN feedback in a cool core cluster, with the 
X-ray surface brightness (top left), the temperature (top right),
the cold gas filaments (bottom left) and metallicity (bottom right),
from Gaspari et al. (2012).}
\label{fig:fig1}
\end{figure}

A large variety of simulations of AGN feedback in clusters or massive elliptical galaxies
have been performed, with different degrees of approximaions and sophistication
(Brueggen et al. 2007, Cattaneo \& Teyssier 2007, Dubois et al. 2010, 
Gaspari et al. 2011, 2012). They vary in the
cooling rate, or the assumption for accretion (boosted Bondi rate, but cold gas accretion 
gives better results). None is self-consistent and can predict the accretion
rate, which depends on subgrid physics. The radiation pressure mechanism
appears insufficient to impact a significant part of the cluster.
Mechanical feedback with jets or winds is required.
The most recent simulations (Gaspari et al. 2012, see Figure 1),  
succeed in moderating the cooling, and keep the cool core  structure,
while previous ones were highly destructive.
The efficiency $\epsilon$, ratio between kinetic energy in the jet and accretion energy, can be scaled to
the structure, $\epsilon$ = 3 10$^{-4}$ for ellipticals and  5 10$^{-3}$ for clusters.

 Are the starbursts related to AGN activity in simulations?
Thacker et al. (2014) have performed a
comparison between the various models in diagrams of star formation rates (SFR)
versus BH accretion rates (BHAR).
During a merger, accreted gas fuels both the SFR and BHAR. A BHAR delay is
expected  since SN feedback is too strong at the beginning (Wild et al. 2010).

\section{Molecular Outflows}

In recent years, the discovery of many massive molecular outflows
has given support to AGN feedback. In the prototypical Mrk 231
both a nuclear starburst and the AGN contribute to the gas outflow,
of $\sim$ 700 M$_\odot$/yr (Feruglio et al. 2010). Some outflows
have been resolved with a size of a few kpc, significant to impact
the galaxy, and provide some quenching. The kinematic power is
of the order of 2 10$^{44}$ erg/s, and the energy of the AGN is required.
High density is traced in the outflows through HCN, HCO$^+$ (Aalto et al. 2012).
Even more powerful outflows are observed at high redshift
(Maiolino et al. 2012).

Cicone et al. (2014) reported interesting relations between
outflow rates and AGN properties, for the molecular outflows known.
For AGN-hosts, the outflow rate correlates with the AGN power
(see Fig 2). The momentum of the outflow is proportional
to the photon momentum and is about 20 L$_{AGN}$/c. 
This is interesting to constrain the physical mechanisms
of the flow. This boosted momentum can be explained by 
energy-driven outflows (Zubovas \& King 2012).

\begin{figure}[ht]
\centering
\includegraphics[width=0.95\columnwidth]{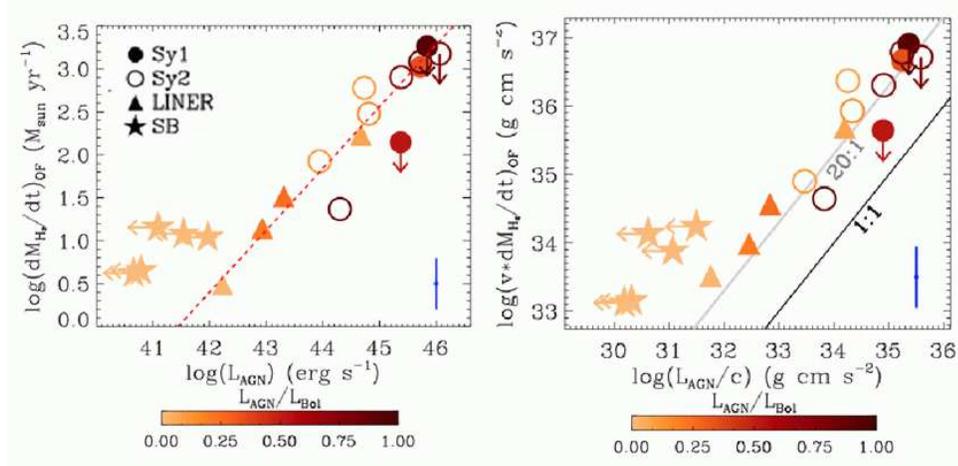} 
\caption{Relation between outflow rate and AGN luminosity (left),
and outflow momentum with  L$_{AGN}$/c, from Cicone et al. (2014).
Colors indicate the  L$_{AGN}$/L$_{bol}$ ratio. The various symbols distinguish
the AGN types, and the stellar symbols are starbursts.}
\label{fig:fig2}
\end{figure}

The question of the very existence of fast molecular outflows is 
still unsolved.
Outflowing gas is accelerated by a shock, and heated to 10$^6$-10$^7$K. 
Molecules should be dissociated at such temperatures.
Even if cold clumps are carried out in the flow, there should
exist some shock signature.
One solution proposed is that
radiative cooling is quick enough to reform molecules in a
large fraction of the outflowing material (Zubovas \& King 2014).
With V$\sim$1000 km/s, and dM/dt $\sim$1000 M$_\odot$/yr, efficient cooling
produces multi-phase media, with triggered star formation.

\section{Mechanisms} 

A very schematic view of the shocks generated by an AGN
wind is displayed in Fig 3 left. The shocked ISM after the discontinuity region
(region c) cools efficiently (free-free emission, metal cooling), and the flow becomes unstable;
if the ratio of the radiation pressure to gas pressure is lower than 0.5,
then the gas collapse in clumps, detaching from the hot flow (Krolik et al. 1981).
The region (c) becomes multiphase, with Rayleigh-Taylor instabilities.
The time-scale for cooling is much lower than 1 Myr, and
star formation is induced. This may explain the tight
correlation between starbursts and AGN.
Also it might then be difficult to disentagnle
supernovae or AGN outflows. All could have been triggered by the AGN
(Zubovas \& King 2014).

\subsection{Energy or momentum conserving outflows}

As shown in Fig 3, if the region (b) of the shocked wind is
cooling efficiently, then energy is not conserved, but only the
momentum, this is the case of a momentum-conserving outflow.

But for very fast winds v$_{in} >$ 10 000 km/s and up to 50 000 km/s, radiative losses are slow,
and almost all the energy can be conserved in this case, called
energy-conserving flow  (Faucher-Gigu\`ere \& Quataert 2012).
Then dM$_{in}$/dt v$_{in}^2 ~\sim$ dM$_s$/dt v$_s^2$, and there
can be a boost in the outflow momentum, 
as large as  v$_{in}$/(2 v$_s)\sim$ 50!  This explains why 
the molecular outflows are observed with a momentum flux $>>$ L$_{AGN}$/c.
It is the push by the hot post-shock gas which boosts the momentum and the velocity
$v_s$ of the swept-up material.
In some cases, even slow winds v$_{in} \sim$ 1000 km/s driven by radiation
 pressure on dust, could be energy-conserving.
The phenomenon is analogous to the
adiabatic phase, or Sedov-Taylor phase in supernovae remnants.
This momentum boost increases the efficiency of supernovae feedback
in galaxies.

\begin{figure}[ht]
\centering
\includegraphics[width=0.95\columnwidth]{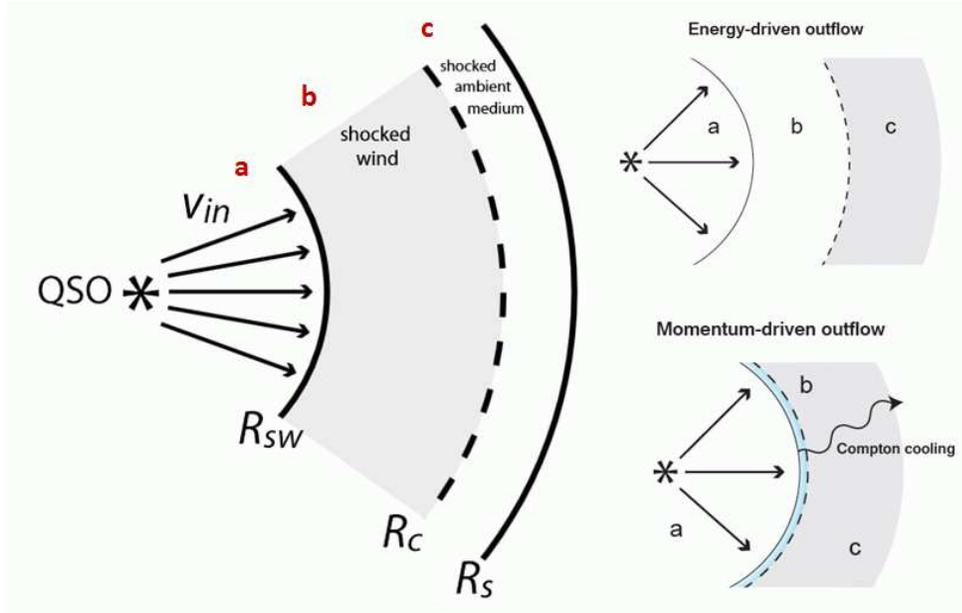} 
\caption{Schematic radial representation of the AGN wind and shock induced
in the surrounding ISM. (a) is the wind region; delimited by a reverse shock (R$_{SW}$)
(b) is the shocked wind, ending at the discontinuity surface (R$_C$, dash line), (c)
is the shocked ISM, bounded by the forward shock (R$_S$), expanding into the 
unperturbed ISM. The two panels at right indicate the cases of
energy-driven outflow (top) where the region (b) is not cooling and energy is conserved there,
and momentum-driven outflow (bottom), where the region (b) cools rapidly, loses its
pressure support, and becomes thin (from Costa et al. 2014).}
\label{fig:fig3}
\end{figure}

Within this framework, it was possible to build
a realistic model and outflow solution for
the typical case of Mrk231, A momentum flux of
15  L$_{AGN}$/c has been obtained
(Faucher-Gigu\`ere \& Quataert 2012).

\subsection{Winds launch}

Towards the very center, ultra-fast outflows (UFO)
 or relativistic winds have been observed in UV absorption lines
(BAL quasars), or from X-rays coronae (Tombesi et al. 2010, 2014).

These winds can be launched from accretion disks,
through several physical mechanisms. 
Thermal heating (Compton) makes the gas reach the escape
velocity. The radiation pressure on electrons (super Eddington regime),
or even radiation pressure on dust, 
or magnetic driving could be the source  (Proga 2003, 2005).
More realistically, 
all driving mechanisms may act together.

Simulations of the {\bf quasar mode}
have been performed taking into account the multi- phase medium
(e.g. Nayakshin 2014).
Most of the outflow kinetic energy 
escapes through the voids.
Cold gas is pushed by ram-pressure, but there is
more feedback on low-density gas.
Both positive and negative feedbacks are observed.
The simulations account for the M-$\sigma$ relation.

Simulations of the {\bf radio mode} have also been performed
with a fractal structure of the gas
(Wagner \& Bicknell 2011).
Relativistic jets produce a very efficient feedback,
and impact on the galaxy, in spite of the porosity of the ISM.

\subsection{Positive AGN feedback} 
 Many simulations reveal signs of positive feedback
(e.g. Silk 2005, Dubois et al. 2013). The phenomenon is
more difficult to observe, but some systems do
show evidence of jet-induces star formation, 
like the Minkowki object, Centaurus A, or
for instance the young, restarted radio loud AGN 4C12.50.
The outflow is located 100 pc from the nucleus
where the radio jet interacts with the ISM
(Morganti et al. 2013, Dasyra \& Combes 2012).

\subsection{Feedback in low-luminosity AGN}

Feedback can also play a role in low-luminosity AGN,
in moderating star formation. The modest impact here is compensated
by the large numbers of such objects. Recently the ALMA observations
at 25 pc resolution of the CO(3-2) line in the barred spiral 
NGC 1433 has revealed a molecular ouflow on the minor axis
(Combes et al. 2013, see Fig 4.).
 About  7\% of the molecular mass detected belongs to the
outflow, with a velocity $\sim$ 100 km/s, uncertain because of 
ill-known inclination. 
This is the smallest flow detected up to now. The kinetic luminosity
is $\sim$  2.3 10$^{40}$ erg/s, and the 
flow momentum is $>$ 10 L$_{AGN}$/c.

The molecular gas is still fueling the AGN, as shown by
the computed gravity torques of the bar (Smajic et al. 2014).

\begin{figure}[ht]
\centering
\includegraphics[width=0.7\columnwidth]{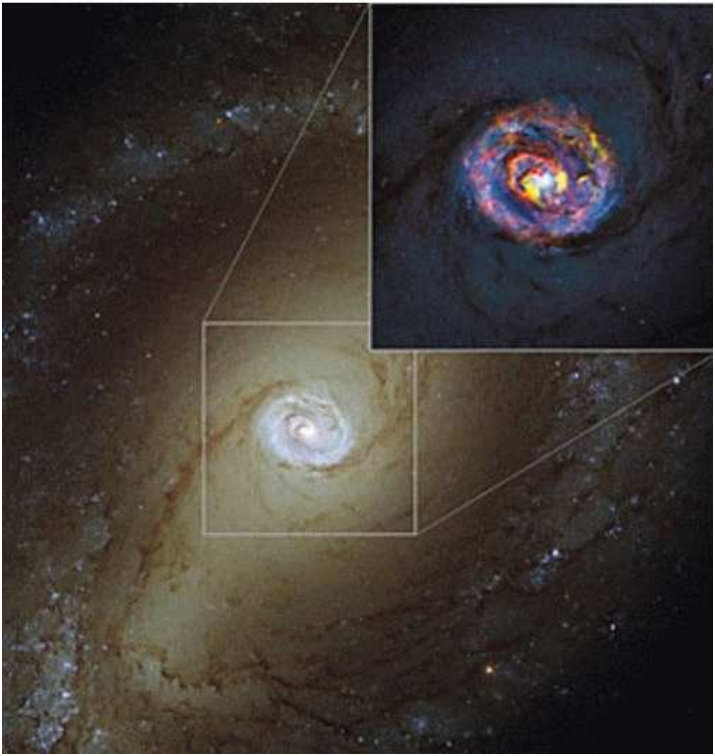} 
\caption{The NGC 1433 barred spiral is a Seyfert 2 of low-luminosity.
The CO(3-2) ALMA map of the nuclear disk shown here superposed on the HST image
has revealed an outflow on the minor axis, the smallest molecular flow detected
up to now (Combes et al. 2013).}
\label{fig:fig4}
\end{figure}

\section{Modes of quenching}

There are two main modes to consider, a
rapid one, with time-scales of 100 Myr, in galaxy mergers,
 with feedback from supernovae and  from AGN, and 
a slow one, with 2-4 Gyr time-scales, either through  morphological 
quenching after bulge formation, or through strangulation, where the
replenishment of the disk in gas is stopped
(Schawinski et al. 2014).
Mild feedback could also delay star formation, in a secular way.
It is however observed that the co-evolution of galaxies and black holes
is no longer tight for low-mass galaxies, governed by secular evolution (e.g. Kormendy \& Ho 2013).

It is not sure whether a feedback loop is required to
explain the M-$\sigma$ relation.
Simulations of torque limited growth of the central BH
have retrieved the relation, without any feedback 
(Angles-Alcazar et al. 2013).
The galaxy and BH grow mostly through gas accretion, sometimes
mergers (but they are not essential).
The numerical evolution has been computed 
with different M-$\sigma$ relations for seeds.
However, the effects of initial conditions are 
quickly erased. The growth of the BH,
dM$_{BH}$/dt$\sim$SFR with scatter, and
no feedback loop is required.

AGB feedback in mergers have been invoked for a long time
for theoretical reasons, and simulated with the usual 
sub-grid physics
(e.g. Springel et al. 2005, Hopkins et al. 2006).
However, the amount of feedback, and how efficiently it couples
with the galaxy disks is unknown. 
Recently, Gabor \& Bournaud (2014) claim that 
the AGN quenching effect on star formation is negligible,
although a significant amount of hot gas is expelled in the intergalactic
medium.

Several modes have been simulated, with more sophisticated details:
{\bf quasar mode}, when dM$_{BH}$/dt $>$ 0.01 Eddington rate, and the energy 
is released spherically; 
{\bf radio-jets} otherwise, with a velocity V= 10$^4$ km/s
imposed  in a cylinder perpendicular to the disk.

The efficiency to form stars can be reduced by a factor 7,
and simulations do show a decrease of the baryon concentration
(Dubois et al. 2013).

Costa et al. (2014) have reproduced the two cases
energy and momentum driven winds, and find
the energy-driven mode much more efficient,
with outflow momenta $>$ 10 L$_{Edd}$/c.
The entrained cold gas has masses $>$ 10$^9$ M$_\odot$
if the after-shock gas cools with metals.

\section{Conclusions}

AGN feedback is very efficient in cool core clusters to moderate
the cooling. The mechanism here is  mechanical with radio jets and cold gas accretion.

 Molecular outflows are now observed frequently around  AGN, with velocities 
v=200-1200 km/s, outflowing masses 10$^7$-10$^9$ M$_\odot$, and load factors $>$3.

The mechanisms range from quasar modes (winds) with powerful AGN,
to radio modes (jets), for more massive galaxies with lower Eddington ratios 
(and lower redshift). The fueling could be either cosmic mass accretion or mergers.

Molecular outflows in AGN-hosts reveal boosted momentum flux of  $\sim$20 L$_{AGN}$/c
This can only be obtained through energy-conserving flows.

\section*{Acknowledgements}
\noindent
Thanks for the organisers, and in particular the chair Bodo Ziegler for such an exciting
and well-organised conference. I acknowledge the European Research Council
for the Advanced Grant Program Number 267399-Momentum.


\begin{thebibliography}{99}

\bibitem[]{}
Aalto S., Garcia-Burillo, S., Muller, S. et al.: 2012, A\&A  537, A44
\bibitem[]{}
Angles-Alcazar D., Ozel, F., Dav\'e, R.: 2013, ApJ 770, 5
\bibitem[]{}
Behroozi, P. S., Wechsler, R. H., Conroy, C.: 2013, ApJ 770, 57
\bibitem[]{}
Brueggen M., Heinz S.,  Roediger E., et al.: 2007, MNRAS 380, L67
\bibitem[]{}
Canning R. E. A., Ryon, J. E., Gallagher, J. S., et al.: 2014, MNRAS in press, arXiv1406.4800
\bibitem[]{}
 Cattaneo A., Teyssier R.: 2007, MNRAS 376, 1547 
\bibitem[]{}
Cicone C., Maiolino, R., Sturm, E. et al.: 2014, A\&A 562, A21
\bibitem[]{}
Combes F., Garcia-Burillo, S., Casasola, V. et al.:  2013, A\&A 558, A124
\bibitem[]{}
Costa T., Sijacki D., Haehnelt M. G.: 2014, MNRAS in press arXiv1406.2691
\bibitem[]{}
Dasyra K., Combes F.: 2012, A\&A 541, L7
\bibitem[]{}
Dubois Y., Devriendt J., Slyz A., Teyssier R.: 2010, MNRAS 409, 985
\bibitem[]{}
Dubois Y., Pichon, C., Devriendt J., et al.: 2013, MNRAS 428, 2885
\bibitem[]{}
Edge A. C., Stewart, G. C., Fabian, A. C.: 1992, MNRAS 258, 177
\bibitem[]{}
Fabian A.C.: 2012, ARAA  50, 455
\bibitem[]{}
Faucher-Gigu\`ere C-A., Quataert E.: 2012, MNRAS 425, 605
\bibitem[]{}
Feruglio C., Maiolino, R., Piconcelli, E. et al.: 2010, A\&A 518, L155 
\bibitem[]{}
Gabor J., Bournaud F.: 2014, MNRAS 441, 1615
\bibitem[]{}
Gaspari M.,  Melioli, C., Brighenti, F., D'Ercole, A.: 2011, MNRAS 411, 349
\bibitem[]{}
Gaspari M., Brighenti, F., Temi, P.: 2012, MNRAS 424, 190
\bibitem[]{}
G\"ultekin, K., Richstone, D. O., Gebhardt, K. et al.: 2009, ApJ 698, 198 
\bibitem[]{}
Hopkins P., Hernquist, L., Cox, T. J. et al.;  2006, ApJS 163, 1
\bibitem[]{}
Kormendy, J., Ho, L. C.: 2013, ARAA 51, 511
\bibitem[]{}
Krolik, J. H., McKee, C. F., Tarter, C. B.: 1981, ApJ 249, 422
\bibitem[]{}
Maiolino R., Gallerani, S., Neri, R. et al.:  2012, MNRAS 425, L66
\bibitem[]{}
McNamara, B. R., Kazemzadeh, F., Rafferty, D. A. et al.: 2009, ApJ 698, 594
\bibitem[]{}
Morganti R., Fogasy, J., Paragi, Z. et al.: 2013, Science 341, 1082
\bibitem[]{}
Nayakshin S.: 2014, MNRAS 437, 2404
\bibitem[]{}
Proga, D. 2003, ApJ 585, 406
\bibitem[]{}
Proga, D. 2005, ApJ 630, L9
\bibitem[]{}
Revaz Y., Combes F., Salom\'e P. 2007, A\&A 477, L33
\bibitem[]{}
Salom\'e P., Combes F., Edge A.  et al. 2006, A\&A 454, 437
\bibitem[]{}
Salom\'e P., Combes F., Revaz Y.  et al. 2008, A\&A 484, 317
\bibitem[]{}
Schawinski K., Urry, C. M., Simmons B. D. et al.: 2014, MNRAS 440, 889
\bibitem[]{}
Silk J.: 2005, MNRAS 364, 1337
\bibitem[]{}
Smajic S., Moser, L., Eckart, A. et al.: 2014, A\&A in press, arXiv1404.6562
\bibitem[]{}
Springel, V., Di Matteo, T., Hernquist, L.: 2005, MNRAS 361, 776
\bibitem[]{}
Thacker R. J., MacMackin, C., Wurster, J., Hobbs, A.: 2014, MNRAS in press, arXiv1407.0685
\bibitem[]{}
Tombesi F., Cappi M., Reeves J.N. et al. : 2010, A\&A 521, A57
\bibitem[]{}
Tombesi F., Tazaki F., Mushotzky R. F. et al.: 2014, MNRAS in press arXiv1406.7252
\bibitem[]{}
Wagner A.Y., Bicknell G. V.: 2011, ApJ 728, 29
\bibitem[]{}
Wild V., Heckman, T., Charlot, S.: 2010, MNRAS 405, 933
\bibitem[]{}
Zubovas K., King A.: 2012, ApJ 745, L34
\bibitem[]{}
Zubovas K., King A.: 2014, MNRAS 439, 400


\end{thebibliography}
\end{document}